\documentstyle[pre,twocolumn,aps,epsfig]{revtex}
\begin{document}
\draft
\title{\bf Self-Organized States in Cellular Automata: Exact Solution}
\author{M.V. Medvedev \cite{mm} and P.H. Diamond \cite{pd}}
\address{Physics Department, University of California at San Diego,
La Jolla, California 92093}
\maketitle

\begin{abstract}
The spatial structure, fluctuations as well as all state probabilities
of self-organized (steady) states of cellular automata can be found (almost)
 exactly and {\em explicitly} from their Markovian dynamics. The method is
shown on an example of a natural sand pile model with a gradient threshold. 
\end{abstract}

Complicated dynamics of various discrete systems may naturally be modeled
by CA having rather simple iteration rules. In particular,
CA models are useful to study traffic jams \cite{traffic},
granular material dynamics \cite{granular}, and self-organization \cite{BTW,Kadanoff}.
The transport properties of systems the, especially in the Self-Organized (SO) 
critical regime were extensively studied 
(cf., \cite{Kadanoff,referee-stat,dynamics,rg-soc,PD-soc}).
However, for many applications, one 
needs to know an entire structure of the SO states. These may include:
(i) the temperature profiles for the convection dominated thermoconduction
and turbulent convection \cite{turb-cond}
(as in the convective zone of the sun and stars),
(ii) stability and average profiles of granular materials (e.g., the sand pile
profiles) and granular flows \cite{granular,BTW}, 
(iii) equilibrium and steady state profiles of plasma pressure and temperature 
in fusion devices \cite{PD-soc,N-Carreras}, etc. 
Despite its importance, the problem of spatial structure and characteristics 
of the SO steady states has received minor attention 
\cite{rg-soc,referee-prof,Dhar,toppl-waves}.
In this Letter, we propose a method which is equally applicable to any CA provided 
its rules are rather simple. We illustrate it on the simplest and popular example 
of an one-dimensional sand pile with a gradient stability criterion. Note that this
model is the closest to a natural pile of sand created by random sprinkling
of sand and with a known repose angle. We focus ourselves on calculating of an 
average slope (the calculation is more transparent) though calculating other
characteristics is trivial.

An interesting result obtained is that the SO profiles of a local slope are 
{\em non-trivial} even for this simplest case.
They are typically flat (e.g., linear) throughout the pile while a narrow region
(similar to  the boundary layer) with rapidly increasing gradient always 
occur near the top of the pile. This picture is very similar to that
experimentally measured in a strongly turbulent convection of a
passive scalar (temperature) with a mean gradient
\cite{turb-cond}. The region of the super-critical (unstable) gradient 
may form near the bottom of the pile when the noise is strong enough
to maintain almost continuous sand flow (overlapped avalanches).
This is in excellent agreement with direct numerical simulations \cite{N-Carreras}.

Among many sand pile models, the Abelian model is the only one that 
was proved to be {\em analytically}
exactly solvable \cite{Dhar,exac-index}.
It is, however, rather
unnatural since its stability criterion is given in terms of a site {\em height},
not {\em slope} of the pile (which depends on neighboring sites, too).
Another tacit assumption is one of {\em weak noise}: no sand is added to the pile
unless at least one unstable site presents (i.e., until the very last 
avalanche is gone). Note that ``waves of toppling''
\cite{toppl-waves}, which are the main point of the whole analysis, 
are well defined in the {\em weak noise} limit, only.
We show that both these restrictions may be relaxed: (i) a site stability
criterion may depend on site's neighbors and (ii) sand may be added
at every time step thus affecting an avalanche dynamics. We also show that
in the weak noise limit, all state probabilities can be calculated 
exactly in a {\em closed form}, while for the strong noise, they can
be found {\em almost} exactly, i.e., with any {\it a priori} given accuracy.

For a general $N$-dimensional sand pile automaton, the procedure is following.
(1) Reformulate the model in the representation where the stability is {\em local}
and defined by the state of a site, alone. This can be done at expense of
introducing a nonlocality to the toppling and noise rules (i.e., they may
depend on states of adjacent sites too). (2) Consider the dynamics of a single 
site. Since toppling rules have no intrinsic memory, however, it is Markovian. 
Construct an $N$-dimensional Markov hyper-lattice (an analog of a Markov chain
in one dimension) with the transition probabilities defined by the CA rules.
All the transition probabilities which depend on states of other sites are,
for now, free parameters. Introducing a generating function, one can then 
solve the problem for a single site. (3) Noting that all sites are identical,
we relate the Markov transition probabilities for different sites. Boundary
conditions then uniquely define their values and, thus, the SO state of the pile.
Note that a mean-field type closure is needed at step (3), only.

{\it Model} ---
The model we consider consists of  $L+1$ spatial {\em sites}, 
numbered from $x=0$ to $x=L$. To each site $x$ is assigned a variable $h(x)$, 
the height of the site. CA rules are applied to the pile at each time-step.
Sand grains are added randomly to sites with probabilities $p_{sand}(x)$ 
increasing the height by one. When the site $x$ is unstable, i.e., if the local 
slope [difference of heights of two neighboring sites $h(x)-h(x+1)$] exceeds 
some critical value $\Delta h_{crit}(x)$, ${N_f}$ sand grains topple onto 
the neighboring site $x+1$ (local, limited model \cite{Kadanoff}). 
Note that ${N_f}>1$ corresponds to the physical situation where friction 
between sand grains at rest is greater than friction of those in motion. 
Sand is expelled from the pile through the right end $x=L$.

For the stability condition to be local, we represent the sand pile in
``gradient space''. We assign to any site the {\em local difference} of 
heights of the nearest neighbors $\Delta h(x)=h(x)-h(x+1)$. 
Then the noise and toppling rules become, respectively:
\begin{mathletters}
\label{grad}
\begin{eqnarray}
\textrm{noise}\ \ & &{\left\{ \begin{array}{ll}
&\Delta h(x-1) \to \Delta h(x-1)-1 ,\\
&\Delta h(x) \to  \Delta h(x)+1 ;
\end{array}\right.}
\label{grad-a}\\
\textrm{toppling}\ \ & &{\left\{ \begin{array}{ll}
&\Delta h(x-1) \to  \Delta h(x-1)+{N_f} ,\\
&\Delta h(x) \to \Delta h(x)-2{N_f} ,\\
&\Delta h(x+1) \to  \Delta h(x+1)+{N_f} .
\end{array}\right.}
\label{grad-b}
\end{eqnarray}
\end{mathletters}
The left end ($x=0$) is now an open boundary (top of the pile):
$\Delta h(0) \to \Delta h(0)-2{N_f}$,  $\Delta h(1) \to  \Delta h(1)+{N_f}$;
the right end ($x=L$) is a closed boundary (bottom of the pile):
$\Delta h(L-1) \to \Delta h(L-1)+{N_f}$, $\Delta h(L) \to  \Delta h(L)-{N_f}$.
Note that ``particles'' in the gradient space are {\em not} sand grains! They
may enter or leave the system through the left boundary, only. Noise creates
a ``particle'' at ${x=0}$ with the probability ${p_{sand}(0)}$, as follows from
Eq. (\ref{grad-a}) for ${x=0}$. A toppling at ${x=0}$ results in a sudden loss
of $N_f$ ``particles''.

{\it Zero-dimensional pile} ---
Now, we may consider one site $x$, alone. It is described by a collection of 
{\em states} representing all possible values of {\em local gradient}. Negative
 states are not allowed. These states are labeled 
by an integer variable $k\equiv\Delta h(x)$, and the critical slope is 
${Z_c}\equiv\Delta h_{crit}(x)$. The states $k<{Z_c}$ are stable, those $k>{Z_c}$ are 
unstable, and the state $k={Z_c}$ is marginally stable.
We introduce the probabilities $p_k$ for a site to occupy a state $k$, i.e., to 
have the slope $\Delta h=k$. Due to noise and overturning events, the  state
of a site will evolve in time. The rules given by 
Eqs.\ (\ref{grad}) are independent of previous history of a system.
Therefore, they define the evolution
of  the slope of a site to be {\em a Markov process}. The states  arranged in
increasing order of $k$ form a Markov chain. Adding and toppling 
rules specify transition probabilities from one state to another on this chain. 

{\bf 1.} {\em Adding sand} [Eq.\ (\ref{grad-a})] results in jumps by 1 right or
left (i.e., a increase or decrease gradient). The transition probabilities of the
process are $\alpha$ and $\beta$, respectively, and equal: 
$\alpha=p_{sand}(x+1)\bigl[1-p_{sand}(x)\bigr],~
\beta=p_{sand}(x)\bigl[1-p_{sand}(x+1)\bigr]$. Note here that {\em adding} of
a sand grain in real space results in {\em increase} or {\em decrease} of a state
(i.e., local slope) in gradient space.

{\bf 2.} {\em Toppling of a site} [Eq.\ (\ref{grad-b})] results in a jump by
$2{N_f}$ states left (i.e., a decrease in gradient). The probability of that 
process is ${\bf 1.~}$, i.e., an unstable state topple on the 
next time-step with the probability unity.

We introduce two ``nonlocal'' transition probabilities.
(1) {\em Toppling of one} of two neighboring sites results in a jump by 
${N_f}$ states right (i.e., an increase in gradient). The probability of this process
is written as $\epsilon^*$.
(2) {\em Toppling of both two} neighbors results in a jump by $2{N_f}$ states
right. The transition probability is written as $\delta^*$.
Both $\epsilon^*$ and $\delta^*$ are simply constants here which are
to be specified in the one-dimensional model via a mean-field-type closure.
Generally speaking, $\alpha,~ \beta,~ \epsilon^*$, and $\delta^*$ will depend on 
$x$ and $k$. We later consider the case where all are independent of 
$x$ and $k$ (homogeneous pile with no local slope dependence). 

\begin{figure}
\epsfig{file=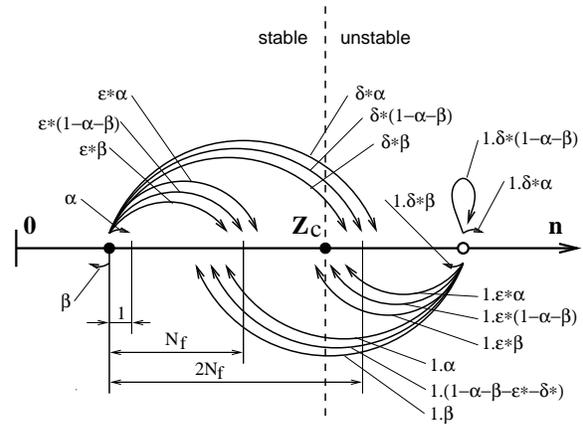,width=3in}\\
\caption{The Markov chain representing a collection of states (slopes) for any
site $x$ in the sand pile model.}
\label{fig1}
\end{figure}

In Fig.\ \ref{fig1}, the Markov chain with all possible transitions for stable 
($\bullet$) and unstable ($\circ$) states with corresponding 
transition probabilities is shown. Noise results in
one-step random walk of a particle on this chain. Toppling of sites results in
jumps by ${N_f}$ and $2{N_f}$ states. Since the noise process is statistically 
independent of toppling, these processes may combine with each other
resulting in jumps by ${N_f}-1, {N_f}+1, 2{N_f}-1$, and $2{N_f}+1$ jumps, with the 
probabilities respectively proportional to $\epsilon^*\beta,~ \epsilon^*\alpha,~
\delta^*\beta$, and $\delta^*\alpha$. All other transition coefficients are 
similarly defined. 

We thus have reduced the problem of a sand pile to the problem of 
{\em a random walk} of a particle on a chain of states where the transition 
probabilities are exactly defined. For a general type of a Markov process
the {\em general kinetic (or master) equation} is
\begin{equation}
\dot p_n(t)=\sum_{k=0}^\infty\left\{\gamma_{nk}p_k(t)-\gamma_{kn}p_n(t)
\right\} ,
\label{gke}
\end{equation} 
where $\gamma_{kn}$ are the transition probability coefficients from state 
$n$ to state $k$. Note that the term $\gamma_{nk}p_k$ describes transitions
{\em into} the state $n$ from state $k$, while $\gamma_{kn}p_n$ 
corresponds to transition {\em out of} $n$ into other states $k$.
This equation defines the probabilities $p_n$ for the  system to be 
in state $n$. The general kinetic equation for one site 
can easily be written using Fig.\ \ref{fig1}. Because of space 
limitations, we do not write  it explicitly.

We introduce {\em a generating function} for the probability distribution $p_k$:
\begin{equation}
F(\zeta,t)=\sum_{k=0}^{\infty}\zeta^kp_k(t) ,
\label{gf}
\end{equation} 
where $\zeta$ can take values $|\zeta|\le 1$ for a series to converge.
The probability distribution can be recovered from the generating function as
\begin{equation}
p_k(t)=(1/k!)\left.d^kF(\zeta,t)/d\zeta^k\right|_{\zeta=0} .
\end{equation} 
Some properties of the generating function are 
${F(1,t)=1},~ {\left.F_{\zeta}'\right|_{\zeta=1}=\left<n(t)\right>}, \dots$, 
where `prime' means derivative.
Here the first is the normalization condition: ${\sum p_j=1}$, and second relates
$F(\zeta,t)$ to the first moment (e.g., expectation value)
of the probability distribution. Higher moments (i.e., standard deviation, etc.)
are obtained from higher derivatives of $F(\zeta,t)$. Multiplying each of 
Eqs.\ (\ref{gke}) for $\dot p_k$ respectively by $\zeta^k$ and taking sum
over all $0\le k<\infty$, we straightforwardly obtain an equation 
for a generating function $F=\Phi+\Psi$. Since we are interesting 
only in a steady state, we set $\dot F=0$. Then it reads
\begin{equation}
F(\zeta)\{\bullet\}
+\Psi(\zeta)\left(\zeta^{-2N_f}-1\right)\!
\left[\alpha^{-1}+\{\bullet\}\right]
=p_0\!\left(\zeta^{-1}-1\right),
\label{gf-full-simp}
\end{equation}
where $\alpha=\beta, \epsilon=\epsilon^*/\alpha, \delta=\delta^*/\alpha$ and  
$
\{\bullet\}=
(\zeta+\zeta^{-1}-2)(1+\epsilon\alpha\zeta^{N_f}+\delta\alpha\zeta^{{2N_f}} )
+\epsilon (\zeta^{N_f}-1)+\delta (\zeta^{{2N_f}}-1).
$
Here 
$\Phi(\zeta)=\sum_{k=0}^{Z_c}\zeta^kp_k$, 
$\Psi(\zeta)=\sum_{k={Z_c}+1}^{\infty}\zeta^kp_k$
are  the partial generating functions of stable and unstable
states, respectively. 

To find $\Psi$, one may use a simple trick [using Eq.\ (\ref{gf-full-simp})]:
\begin{mathletters}
\begin{equation}
\left.\left(\frac{\partial^{Z_c+1+j}}{\partial\zeta^{Z_c+1+j}}\ \Phi(\zeta)
\right)\right|_{\zeta=0}\equiv0 , \qquad j\ge1 .
\label{trick}
\end{equation}
In general, this system is an infinite set of related equations. It relates 
all $p_{Z_c+1+j}$ to $p_0$. Together with ${F(1)=1}$, it provides an {\em exact} 
solution, $F(\zeta)$, of Eq.\ (\ref{gf-full-simp}). For an {\em Abelian} sand pile,
the transition probabilities of simultaneous toppling and noise [e.g., 
toppling to {\em higher} states, ${\bf 1.}\delta^*\alpha$ (see Fig.\ \ref{fig1})]
vanish identically by definition.
(Note, this is the limit when noise is too {\em weak} to affect avalanche dynamics.)
Thus, the highest achievable state is $Z_c+2N_f$. The Markov chain is finite, and
Eqs.\ (\ref{trick}) reduce (schematically) to the system with triangular matrix 
(with $a_{ij}$ and $b_k$ being constants)
\begin{equation}
\pmatrix{
a_{1,1} \!\!\!\!& a_{1,2} & \!\! \dots \!\! & a_{1,2N_{\! f}} \cr
0 \!\!\!\!& a_{2,2} & \!\! \dots \!\! & a_{2,2N_{\! f}} \cr
\vdots \!\!\!\!& \vdots & \!\! \ddots \!\! & \vdots \cr
0 \!\!\!\!& 0 & \!\! \dots \!\! & a_{2N_{\! f},2N_{\! f}} }
\pmatrix{
p_{Z_c+1} \cr
p_{Z_c+2} \cr
\vdots \cr
p_{Z_c+2N_{\! f}} }
\!=\! \pmatrix{
b_1 \cr
b_2 \cr
\vdots \cr
b_{2N_{\! f}} } \! p_0
\label{matrix}
\end{equation}
\end{mathletters}
which can be solved {\em exactly} and {\em explicitly}. In the opposite case,
when noise is {\em not} weak, one may, however, truncate the system of 
Eqs.\ (\ref{trick})  to a finite
hierarchy simply by noticing that the probabilities $p_{Z_c+2N_f+j}$, $j\ge1$ are
very low since they can be reached only from {\em unstable} states. If one
truncates at the state $Z_c+2N_f+j$, the error in determination of all the
state probabilities, $p_k$, will not exceed the value $({\bf 1.}\delta^*\alpha)^j$.

Eqs. (\ref{trick}) or (\ref{matrix}) constitute the (almost) exact solution,
i.e., state probabilities of a site $x$, of a CA sand pile model in terms of state
probabilities (entering through $\epsilon$ and $\delta$) of its neighbours.

The normalization condition, ${F(1)=1}$, gives the relation for
the total probability for a site to be unstable:
\begin{equation}
{\cal P}\equiv\Psi(1)=\alpha[p_0+{N_f}(\epsilon+2\delta)]/2 {N_f} .
\label{P}
\end{equation}
We straightforwardly define the SO profile as the mathematical expectation 
value (average) of the random process:
\begin{equation}
\Delta h_{so}\equiv\langle n\rangle=F_{\zeta}'(1),
\label{soc}
\end{equation} 
where two unknowns $p_0(\epsilon,\delta)$ and $\Psi_\zeta'(1)$ appear and 
are to be found from Eqs.\ (\ref{trick}) or (\ref{matrix}). 
For arbitrary $Z_c$ and $N_f$, (especially for large $Z_c$ and $N_f$, i.e., 
in the continuous limit) the result can be easily found numerically. 
To obtain an analytically tractable expression, 
we make additional (though natural) approximations.

{\bf 1.}~
Let's consider an asymmetric random walk on a finite chain with transition 
probabilities to the right and to the left, $g$ and $r$, respectively. One can 
easily show [recursively from Eq.\ (\ref{gke})] that
$p_0\!\left(\frac{g}{r}\right)=
\left({g\over r}-1\right)\left[\left({g\over r}\right)^{c}-1\right]^{-1}$,
where $c$ is a constant which is found from an expansion of 
$p_0(g/r)$ near $g/r\sim1$ 
to give $p_0|_{g/r=1}=1/c$. By analogy with an asymmetric random
walk, we write $g/r=(\alpha+{N_f}\epsilon^*+2{N_f}\delta^*)/\beta=
1+{N_f}\epsilon+2{N_f}\delta$ and
\begin{equation}
p_0\simeq {N_f}(\epsilon+2\delta)/
\left( [1+{N_f}(\epsilon+2\delta)]^{1/p_0^{(0)}-1} \right) ,
\label{p0}
\end{equation}
where $p_0^{(0)}=p_0|_{\epsilon=\delta=0}=({Z_c}-{N_f}+3/2)^{-1}$ [the last follows 
from Eq.\ (\ref{gf-full-simp}) for $\epsilon=\delta=0$].

{\bf 2.}~
To define $\Psi_\zeta'(1)$, we consider two limits for which $\Psi(\zeta)$ is
known {\em a propri}.
When ${\epsilon=\delta=0}$, only one-step transitions (noise) exist. Therefore,
from the definition, we have 
$\Psi_\zeta'(1)|_{\epsilon=\delta=0}=({Z_c}+1)p_{{Z_c}+1}$,
i.e., only the first unstable state can be achieved. For $\epsilon, \delta$
sufficiently large, the states $k\in [Z_c+1,\ Z_c+2N_f]$ are roughly uniformly
populated while higher states ${Z_c}+{2N_f}+k, k\ge1$ have low probability,
as they can be reached only from {\em unstable} states. Thus, we can write
$\Psi_\zeta'(1)\simeq\Psi(1)\left({Z_c}+1+({2N_f}-1)/2\right)$.
From comparison of the last two equations, we conclude
\begin{equation}
\Psi_\zeta'(1)\simeq{\cal P}\left[({Z_c}+1)+(N_f-1/2)f(\epsilon,\delta)\right] ,
\end{equation}
where $f=1$ for large $\epsilon, \delta$ and vanishes for ${\epsilon=\delta=0}$. 
Because the quantity $\epsilon+2\delta$ is a ``measure of asymmetry''
of a random walk, we choose 
$f(\epsilon,\delta)=2(\epsilon+2\delta)/\left(1+(\epsilon+2\delta)\right)$.

Finally, the SO local slope (for $\alpha\ll1$) reads
\begin{eqnarray}
\Delta h_{so}
&\simeq&\left[p_0+{N_f}(\epsilon+2\delta)\right]\left(
\frac{{N_f}-1/2}{1+{N_f}(\epsilon+2\delta)}+\alpha\right)\nonumber\\
& & { } +\left({Z_c}+1-3{N_f}/2\right)+\delta N_f/(\epsilon+2\delta) .
\label{soc-full}
\end{eqnarray}
Here $p_0$ is given by Eq.\ (\ref{p0}). Eq.\ (\ref{soc-full}) depends on the noise 
strength $\alpha$ as well as on the toppling probabilities of adjacent sites of the pile.

\begin{figure}
\epsfig{file=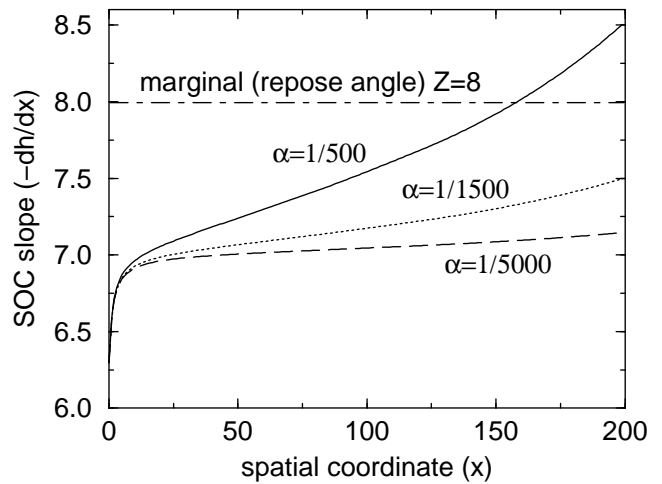,width=3in}\\~ \\
\caption{The SO profiles of gradient of the pile (${Z_c}=8,\ {N_f}=3$) 
for three noise levels.}
\label{fig2}
\end{figure}

{\it One-dimensional pile} ---
Eq.\ (\ref{soc-full}) defines the average SO slope for every site $x$. 
The quantities
$\epsilon$ and $\delta$ are defined by toppling probabilities of neighboring
sites. Each site $x$ topples with probability ${\cal P}={\cal P}(x)$. This 
probability varies from site to site. In the mean-field approximation, by definition
\begin{equation}
\begin{array}{rl}
\epsilon\alpha&={\cal P}(x-1)\bigl[1-{\cal P}(x+1)\bigr]\\
&\qquad +{\cal P}(x+1)\bigl[1-{\cal P}(x-1)\bigr] ,\\
\delta\alpha&={\cal P}(x-1){\cal P}(x+1) . 
\end{array}
\label{e,d}
\end{equation}
Note, the stonger the noise, the better this anzatz works, because of decorrelation
of ${\cal P}(x\pm1)$ caused by noise. 
Eq.\ (\ref{P}) can be written as a recurrence equation for probabilities
${\cal P}(x)$
\begin{equation}
2{\cal P}(x)=\alpha p_0(x)/N_f+\left[{\cal P}(x-1)+{\cal P}(x+1)\right] ,
\label{recurr}
\end{equation}
where $p_0(x)$ is also a function ${\cal P}$ and given by Eq.\ (\ref{p0}).
This equation can be solved numerically with the condition at the open
(left) boundary that ``influx''=``outflux''. The initial value is thus, 
${\cal P}(0)=p_{sand}(0)/{N_f}$. Eq.\ (\ref{soc-full}) together with Eqs.\ (\ref{e,d})
defines a spatial profile of the SO slope of the sand pile. In the continuous limit
(vanishing cell size of a Markov chain), Eq.\ (\ref{recurr}) is equivalent to
${\cal P}''_{xx}=\alpha p_0/N_f$. Thus the approximate solution matching the
boundary condition is
\begin{equation}
{\cal P}(x)\simeq\left({p_{sand}p_0}/{N_f}\right)x^2
+\left({p_{sand}(1-p_0)}/{N_f}\right)x .
\end{equation}
The SO gradient profiles are shown in Fig.\ \ref{fig2} for ${Z_c}=8, {N_f}=3$, 
and three values of noise strength $\alpha\simeq p_{sand}$:\
$\alpha=1/5000$ (low noise), $\alpha=1/1500$ and $\alpha=1/500$ (high noise). 
The average gradient profiles 
always have a region of relatively small gradient, {\em ``boundary layer''},
near the top of the pile. This region appears due to the effect of the open
(in gradient space) boundary at $x=0$. In the case of low noise, the SO
gradient is almost constant throughout the pile and always below the 
marginally stable value. In the ``over-driven'' regime (i.e., high noise), a region
of a {\em super-critical} gradient appears near the bottom of the pile. In this case 
the sand influx is so large that a nearly constant flow of sand forms near the 
bottom, thus maintaining an unstable, super-critical gradient. These results 
agree well with simulations \cite{N-Carreras}.

In this Letter, we show that Abelian property is not necessary for an (almost)
exact solvability of a sand pile CA. As an example, a spatial profile of an 
one-dimensional sand pile is calculated. 

We are grateful to M.N. Rosenbluth for numerous interesting and fruitful
discussions and critical comments.
We also thanks B.A. Carreras, D.E. Newman, and T.S. Hahm 
for discussions. This work was supported by Department of Energy grant 
No. DE-FG03-88-ER53275.

%
\vskip.5in
\end{document}